\documentclass[11pt]{article}

\usepackage[margin=1in]{geometry}
\usepackage{times}
\usepackage[T1]{fontenc}
\usepackage[utf8]{inputenc}
\usepackage{graphicx}
\usepackage{booktabs}
\usepackage{array}
\usepackage{hyperref}
\usepackage{authblk}
\usepackage{caption}
\usepackage{setspace}
\usepackage{enumitem}

\hypersetup{
    colorlinks=true,
    linkcolor=black,
    citecolor=black,
    urlcolor=blue
}

\title{\textbf{Trust by Design: Trust Calibration Through Non-Advisory Socratic Dialogue in Conversational Agents}}

\author[1]{Roba Hassan}
\author[1]{Nahla Aboromi}
\author[1]{Naomi Unkelos-Shpigel}
\affil[1]{Braude College of Engineering, Israel}

\date{}

\begin{document}

\maketitle

\begin{center}
\begin{minipage}{0.9\textwidth}
\small
\noindent\footnotesize Roba Hassan, Nahla Aboromi, and Naomi Unkelos-Shpigel contributed equally to this work. \\
\noindent Roba Hassan: \texttt{Roba.Hassan@e.braude.ac.il} \quad
Nahla Aboromi: \texttt{Nahla.Aboromi@e.braude.ac.il} \quad
Naomi Unkelos-Shpigel: \texttt{Naomius@braude.ac.il}
\end{minipage}
\end{center}

\begin{abstract}
\noindent As conversational AI systems increasingly operate in sensitive domains, the central challenge shifts from usability to trust calibration, ensuring that users rely on systems neither too much nor too little. Systems that provide advice or interpretations risk encouraging inappropriate reliance, particularly when users perceive AI outputs as authoritative. We present CASELy, a conversational agent explicitly designed to limit its own authority through non-advisory Socratic dialogue. The agent asks reflective questions grounded exclusively in user input and refuses to provide advice, recommendations, or interpretations. This design operationalizes trust calibration by constraining agent agency rather than optimizing capability. In a pilot randomized controlled study with higher education students, participants interacting with the Socratic dialogue reported substantially higher user experience (UEQ-S overall = 1.50) compared to a non-dialogue control (0). Qualitative findings identify three mechanisms supporting calibrated trust: transparency through visible grounding, preservation of user decision authority, and reduced fear of judgment. We argue that appropriate reliance can be achieved through interactional constraints, offering a design pattern for trustworthy conversational AI in sensitive contexts.
\end{abstract}


\section{Introduction}

Social-emotional learning (SEL) is increasingly recognized as a critical component of student success in higher education, influencing well-being, engagement, and academic outcomes \cite{conley2015sel, conleydonahue2024sel, simion2023sel}. Empirical studies and reviews consistently demonstrate that SEL competencies such as self-awareness, emotional regulation, and responsible decision-making are associated with improved academic performance and resilience \cite{conley2015sel, simion2023sel, takizawa2024sel}. However, despite this evidence, access to scalable, non-judgmental SEL support remains limited, and many existing programs rely heavily on instructor facilitation or institutional resources \cite{feisal2025sel, desensi2024sel}.

Trust calibration theory emphasizes that trustworthy systems align user trust with system capabilities and intent, rather than maximizing trust \cite{wischnewski2023trust}. When conversational agents provide advice, interpretations, or confident recommendations, users may develop miscalibrated trust, leading to inappropriate reliance or reduced personal agency \cite{wischnewski2023trust, zhang2020confidence}. Research shows explanation and confidence cues can improve perceived trust while simultaneously degrading calibration accuracy \cite{zhang2020confidence}.

Conversational AI systems are often proposed as scalable tools for supporting reflection and learning. Yet, research on human--AI interaction highlights a key risk: when systems provide advice, interpretations, or confident recommendations, users may develop miscalibrated trust, leading to inappropriate reliance or reduced personal agency \cite{wischnewski2023trust, zhang2020confidence}. Trust calibration theory emphasizes that trustworthy systems are not those that maximize user trust, but those that align user trust with system capabilities and intent \cite{wischnewski2023trust}.

In sensitive domains such as SEL, this alignment is particularly important. Pedagogical conversational agents that appear authoritative or evaluative may unintentionally undermine reflective learning by shifting responsibility from the learner to the system \cite{wolfel2023pedagogical}. This creates a tension between providing support and preserving autonomy.

In this work, we explore a design approach that addresses this tension by constraining system behavior rather than enhancing system intelligence. We investigate whether a conversational AI system that explicitly avoids advice, judgment, and interpretation using only Socratic questioning can support reflection while fostering calibrated trust. Our research question is:

\begin{quote}
\emph{How does non-advisory Socratic dialogue in conversational AI influence trust calibration and appropriate reliance in social-emotional learning contexts?}
\end{quote}

\section{The CASELy System}

CASELy is an anonymous, web-based conversational system designed to support reflective SEL while maintaining calibrated trust. The system comprises three stages:

\begin{enumerate}[label=\textbf{\arabic*.}]
\item \textbf{Socio-Emotional Simulation:} Users respond to realistic academic scenarios through open-ended written reflection, designed to activate CASEL competencies.
\item \textbf{Structured SEL Analysis:} User responses are analyzed across five CASEL dimensions, producing descriptive feedback and scores without evaluative or prescriptive language.
\item \textbf{Non-Advisory Socratic Dialogue (Experimental Condition):} The system engages users through open-ended questions grounded strictly in their own responses. It avoids advice, recommendations, or interpretations; asks one question per turn; and concludes with a reflective summary mirroring user language.
\end{enumerate}

A control condition completes the same flow without dialogue. The system follows privacy-by-design principles \cite{vanrest2012privacy}, including full anonymity via randomly generated identifiers and separation between research and user data.

\subsection{Trust Calibration Through Interaction Design}

CASELy operationalizes trust calibration by minimizing interactional signals that inflate perceived system authority. In contrast to advisory pedagogical agents \cite{wolfel2023pedagogical}, the system avoids evaluative or instructional language, grounds all dialogue explicitly in user input, restricts conversational turns to reflection rather than guidance, and makes system limitations continuously visible through behavior.

\begin{figure}[htbp]
\centering
\includegraphics[width=0.85\textwidth]{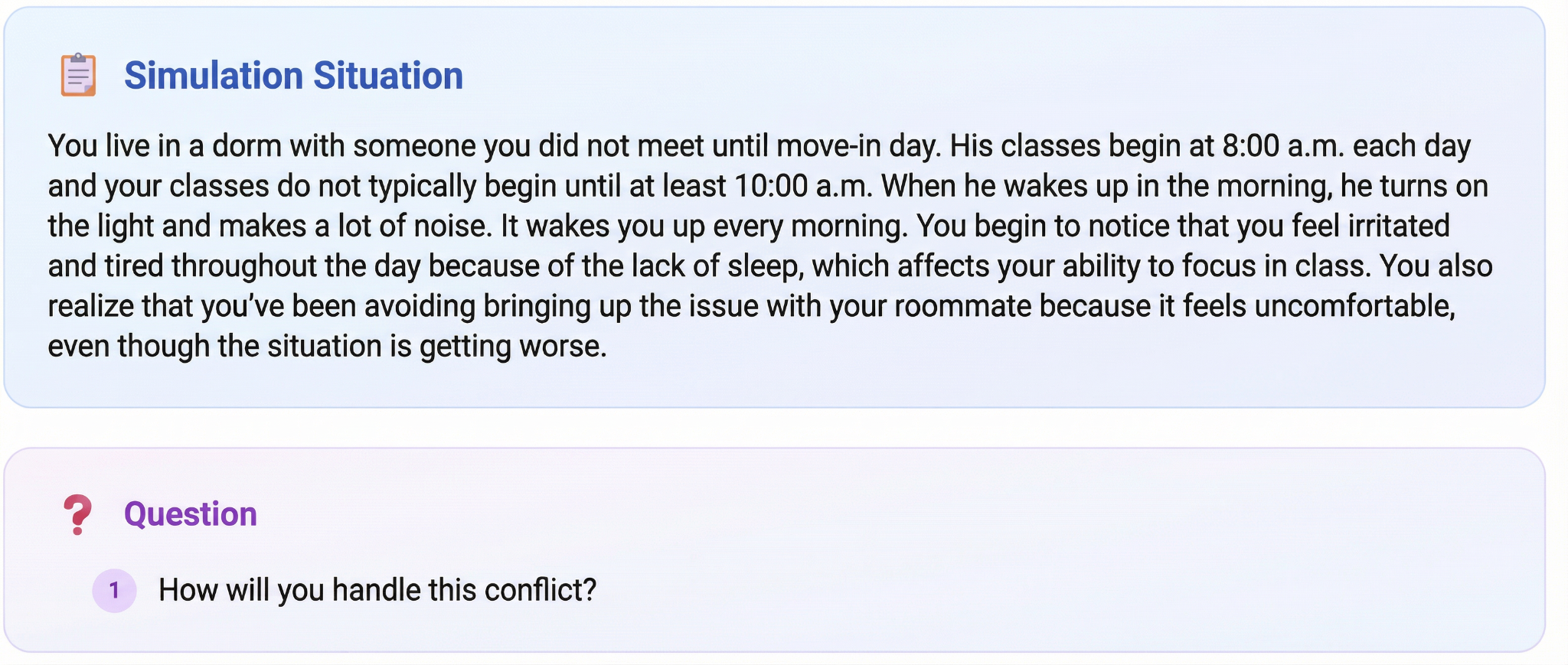}
\caption{Example of a Socio-Emotional Simulation Screen.}
\label{fig:sim}
\end{figure}

This approach aligns with trust calibration theory by ensuring users perceive the system as a reflective aid rather than an expert or decision-maker \cite{wischnewski2023trust}. Instead of increasing trust through confidence or explanation, CASELy aims to prevent over-trust by design. Figure~\ref{fig:sim} presents the socio-emotional simulation screen, where participants engage with a realistic academic scenario and provide an open-ended written response reflecting their emotional and cognitive processing.

\begin{figure}[htbp]
\centering
\includegraphics[width=0.95\textwidth]{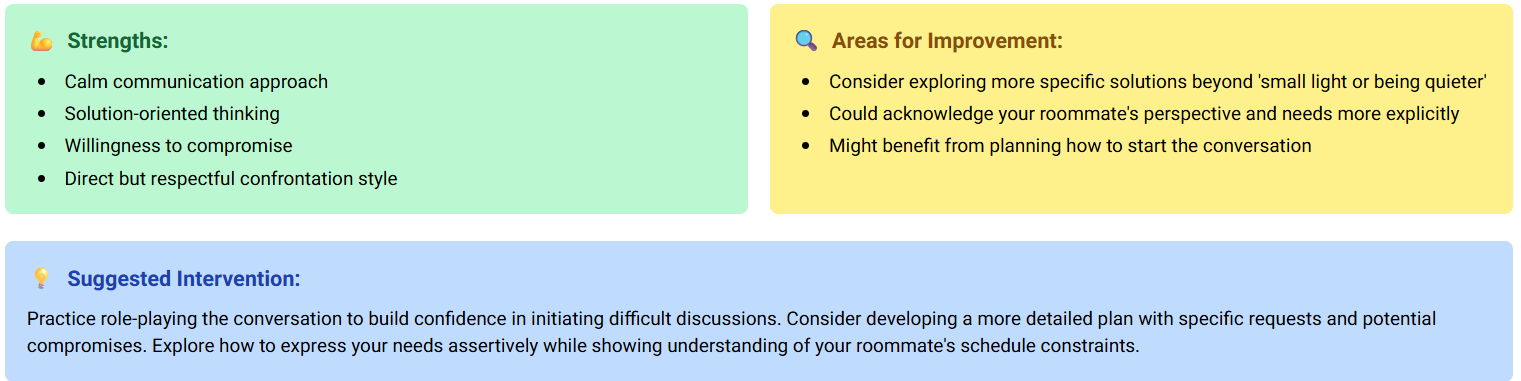}
\caption{Example of AI-Generated CASEL Competency Analysis and Feedback.}
\label{fig:analysis}
\end{figure}

Figure~\ref{fig:analysis} presents automated CASEL analysis with feedback, explanations, and informing stages. The feedback is organized into three components that operationalize trust calibration: Strengths (green) identifies observed competencies grounded in actual user responses; Areas for Improvement (yellow) presents growth opportunities using reflective prompts rather than prescriptive advice; and Suggested Intervention (blue) offers pathways for self-directed learning through practice exercises. This three-part structure implements trust calibration principles by making analysis transparent and verifiable, using descriptive rather than evaluative language, and preserving user autonomy---users can verify feedback against their own responses and maintain decision-making control over their learning trajectory.

\section{Pilot Study and Preliminary Evaluation}

A pilot study involved four senior undergraduate students (three experimental, one control; fourth-year software engineering students, ages 20--25, three females and one male). Participants interacted with the system for approximately 15 minutes and completed the UEQ-S user experience questionnaire \cite{schrepp2017ueqs} and semi-structured interviews.

\begin{figure}[htbp]
\centering
\includegraphics[width=0.9\textwidth]{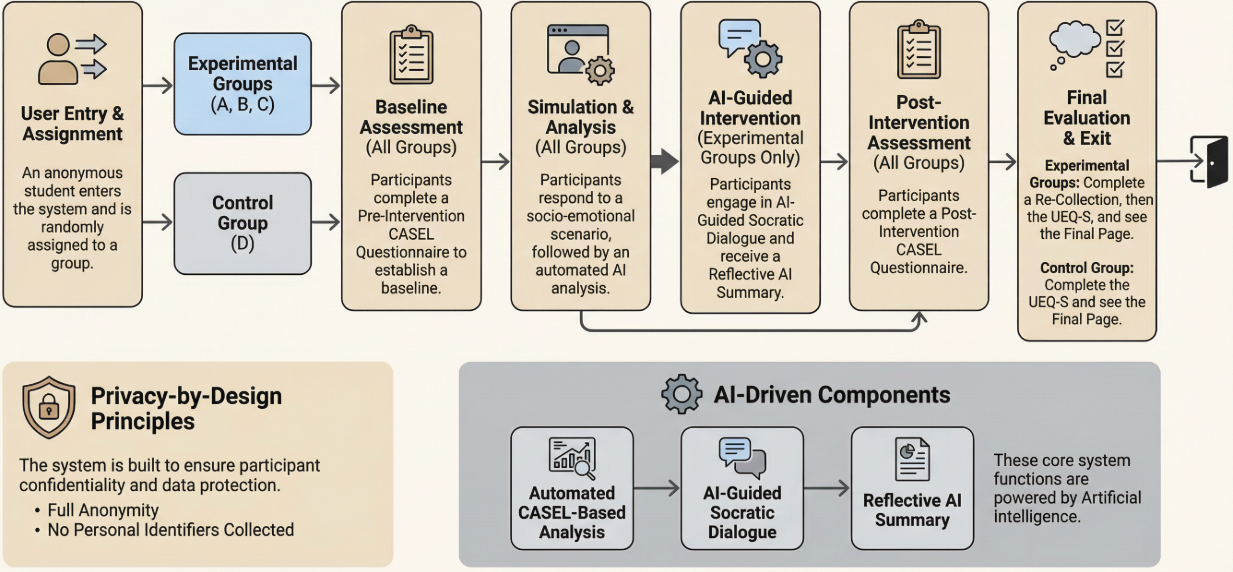}
\caption{Evaluation design.}
\label{fig:design}
\end{figure}

Figure~\ref{fig:design} presents CASELy's experimental architecture, demonstrating how trust calibration principles are operationalized through system design. The flow comprises seven sequential stages: (1) User Assignment randomly distributes participants to three experimental groups (A, B, C) or a control group (D); (2) Baseline Assessment establishes pre-intervention CASEL competencies; (3) Simulation \& Analysis presents realistic scenarios followed by AI-driven descriptive feedback without evaluative judgment; (4) AI-Guided Socratic Dialogue (experimental groups only) delivers non-advisory questions grounded exclusively in user responses---the key intervention absent in control; (5) Post-Intervention Assessment measures changes via the CASEL questionnaire and UEQ-S; (6) Final Evaluation completes the interaction. Two foundational principles support trust calibration: Privacy-by-Design \cite{vanrest2012privacy} ensures full anonymity and psychological safety for honest self-disclosure, while AI-Driven Components (automated CASEL analysis, Socratic dialogue, reflective summary) constrain system authority by analyzing and questioning without advising. This architecture embodies trust calibration theory by isolating the effect of non-advisory interaction patterns---comparing groups with and without dialogue tests whether strategic limitations (refusing advice, visible grounding) enhance appropriate reliance, demonstrating that trust calibration emerges from interaction design rather than capability maximization.

\section{Discussion and Conclusions}

\begin{figure}[htbp]
\centering
\includegraphics[width=0.55\textwidth]{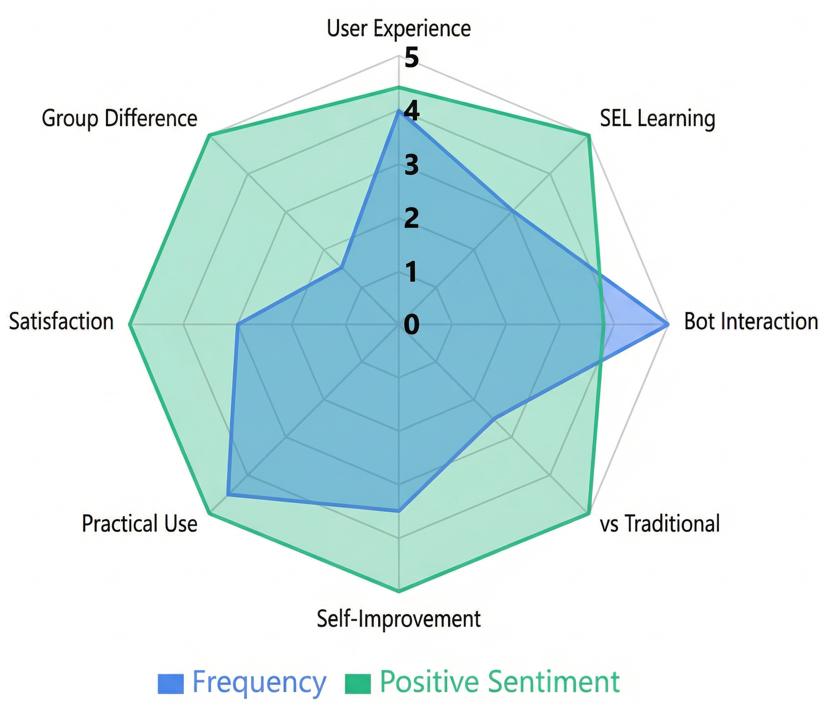}
\caption{Thematic analysis of overall pilot study interviews: frequency and sentiment across categories.}
\label{fig:themes}
\end{figure}

We address our research question through integrated quantitative and qualitative analysis. Quantitatively, participants engaging with Socratic dialogue demonstrated substantially higher UEQ-S scores (overall = 1.50) versus non-dialogue control (0), revealing enhanced user engagement and experience. Qualitatively, Figure~\ref{fig:themes} displays thematic analysis of pilot interviews using a radial 0--5 scale, where the outer axis represents mention frequency and color indicates positive sentiment. Bot Interaction emerged as the most frequently discussed category, while User Experience and SEL Learning received the highest positive sentiment ratings. The divergence between dialogue and non-dialogue conditions is evident across multiple dimensions, with particularly strong positive sentiment for User Experience, Group Difference, and Practical Use. This pattern underscores the centrality of Socratic dialogue in shaping participant experiences, as the interactive component dominated both the frequency and emotional valence of participant reflections.

\textbf{Qualitative Findings.} Thematic analysis of semi-structured interviews following the pilot revealed systematic patterns linking participant experiences to trust calibration theory principles. Three core mechanisms emerged as central to appropriate reliance: preservation of autonomy and decision ownership, reduced fear of judgment in sensitive reflection, and increased trust through visible grounding and limited system authority. Participants consistently recognized and valued the system's constraints---explicitly noting how the refusal to provide advice, the grounding in their own language, and the non-judgmental questioning style shaped their experience. These observations directly validate the theoretical claim that trust calibration can be operationalized through interaction design rather than capability enhancement.

Table~\ref{tab:mapping} systematically maps participant evidence to specific trust calibration principles from the literature, demonstrating how CASELy's design features translate theoretical constructs into observable user experiences. Each row illustrates the alignment between a trust calibration principle, the corresponding system design feature, direct participant testimony, and the theoretical mechanism by which appropriate reliance is achieved.

\begin{table}[htbp]
\centering
\caption{Mapping Participant Experiences to Trust Calibration Principles}
\label{tab:mapping}
\small
\begin{tabular}{p{2.3cm}p{2.6cm}p{4cm}p{3.3cm}}
\toprule
\textbf{Trust Calibration Principle} & \textbf{CASELy Design Feature} & \textbf{Participant Evidence} & \textbf{Theory Alignment} \\
\midrule
Appropriate Reliance \cite{wischnewski2023trust} & Non-advisory dialogue; explicit refusal to provide recommendations & ``At the end [the bot] said to me, `your decision, I respect that.' I appreciate that'' (P2) & Prevents over-reliance by maintaining user as decision-maker rather than delegating authority to system \\
\addlinespace
Transparency Through Constraints \cite{wischnewski2023trust, zhang2020confidence} & Visible grounding in user input; one question per turn & ``It showed that it really analyzes my answers and gives me responses accordingly'' (P1); ``It's not just randomly throwing answers'' (P2) & Makes system capabilities and limitations continuously visible through interaction patterns rather than post-hoc explanations \\
\addlinespace
Avoiding Confidence Inflation \cite{zhang2020confidence} & Honest, non-flattering responses; no evaluative praise & ``If you think correctly, or don't think correctly, he tells you that you think correctly'' (P2) & Avoids artificial confidence signals that can improve perceived trust while degrading calibration accuracy \\
\addlinespace
Preserving User Agency \cite{wolfel2023pedagogical} & Questions rather than advice; mirrors user language & ``An experience rather than just learning\ldots you're talking with someone real'' (P1) & Maintains learner responsibility and active engagement \\
\addlinespace
Emotional Safety for Honest Interaction \cite{vanrest2012privacy} & Anonymity; non-judgmental questioning & ``Sometimes people are embarrassed or get confused when there's someone talking and sitting in front of them'' (P1) & Reduces social desirability bias and enables authentic self-disclosure necessary for trust assessment \\
\addlinespace
Distinguishing System from Human Authority \cite{wolfel2023pedagogical} & Acknowledges limitations; refuses interpretation & ``It's a bot, but you're talking with someone\ldots real'' (P1) & Users recognize system boundaries while still finding interaction valuable---appropriate trust rather than anthropomorphic over-trust \\
\addlinespace
Functional Transparency \cite{wischnewski2023trust} & Structured, organized responses grounded in user input & ``The questions he asked\ldots it was very organized. It showed that he really analyzes my answers'' (P1) & System behavior itself communicates capabilities, creating emergent transparency without requiring model interpretability \\
\addlinespace
Comparative Calibration \cite{wischnewski2023trust, zhang2020confidence} & Interactive engagement vs.\ passive learning & ``The experience of learning through a bot is better, because it's interactive'' (P3); ``I can ask questions, and it answers me'' (P4) & Users appropriately value system for its actual strengths (interactivity, personalization) rather than perceived omniscience \\
\bottomrule
\end{tabular}
\end{table}

These findings align with prior work, suggesting that reduced system assertiveness can improve trust calibration and user experience in automated systems \cite{wischnewski2023trust, zhang2020confidence}. Experimental participants particularly valued the AI-guided dialogue for reducing fear of judgment and enabling honest self-expression.

This work contributes to research on trustworthy conversational AI by demonstrating that trust calibration can be achieved through deliberate interactional constraint. In the context of social-emotional learning, non-advisory Socratic dialogue supports reflective engagement while avoiding risks associated with over-reliance and misplaced authority.

Rather than treating trust as a quantity to be maximized, CASELy aligns with contemporary perspectives that frame trust as a relationship to be calibrated \cite{wischnewski2023trust}. We argue that conversational AI systems in sensitive educational contexts should prioritize appropriate reliance over perceived intelligence or helpfulness.

By combining quantitative and qualitative evaluation within a controlled framework, CASELy offers a scalable, replicable, and ethically grounded model for AI-mediated SEL interventions that support reflective learning without compromising autonomy or privacy.

Future iterations will evaluate CASELy with larger samples, explore longitudinal trust dynamics, and examine transferability to other sensitive domains requiring autonomy-preserving support.

\bibliographystyle{unsrt}
\bibliography{references}

\end{document}